\documentstyle[aps,epsf,prl,multicol]{revtex}

\begin{document}
\draft
\title{Phase slips and phase synchronization of coupled oscillators}

\author{Zhigang Zheng$^{a,b}$ Gang Hu$^{a,b}$ and Bambi Hu$^{a,c}$}
\address{
$^a$Department of Physics and Center for Nonlinear Studies, Hong Kong
Baptist University, Hong Kong \\
$^b$Department of Physics, Beijing Normal University, Beijing 100875, 
China \\
$^c$Department of Physics, University of Houston, Houston TX77204, USA}
\maketitle

\begin{abstract}
The behaviors of coupled oscillators, each of which has periodic motion with
random natural frequency in the absence of coupling, are investigated. 
Some
novel collective phenomena are revealed. At the onset of instability of the
phase-locking state, simultaneous phase slips of all oscillators and 
quantized phase shifts in these phase slips are observed.
By increasing the
coupling, a bifurcation tree from high-dimensional quasiperiodicity to chaos
to quasiperiodicity and periodicity is found.
Different orders of phase synchronizations of chaotic oscillators and
chaotic clusters play the key role for constructing this tree structure.
\end{abstract}

\begin{multicols}{2}
The investigation of coupled oscillators has attracted constant interest for
many decades [1]. The rich collective behaviors of these systems, such as
mutual entrainment, self-synchronization, and so on, are observed in many
fields, e.g., coupled laser systems, Josephson junction arrays, biological
and chemical oscillators etc. [2-10]. In early studies, interest was focused
on coupled oscillators, of which each is periodic without coupling.
Recently, the investigation has been extended to coupled chaotic systems
(i.e., individual systems are chaotic without coupling). Significant
phenomena were found, such as phase synchronization of two mutually coupled
chaotic oscillators [11, 12], and clustering and cluster-cluster synchronization
of multiple coupled chaotic units for local [13] and global [14] couplings.

In this letter, we study the following $N$ coupled oscillators with
the nearest coupling,
\begin{equation}
\label{1}\stackrel{.}{\theta }_i=\omega _i+\frac K3[\sin (\theta
_{i+1}-\theta _i)+\sin (\theta _{i-1}-\theta _i)], 
\end{equation}
$i=1,2,...,N$, where $K$, $\theta _i$ and $\omega _i$ are the coupling
strength, the angle of modulo $2\pi $ and the natural frequency of the
$i$-th oscillator, respectively. Model (1) has been extensively investigated
in the past several decades. Here we concentrate on the dynamical behavior
of the system. In particular, we are interested in the characteristic features
of the motions of individual oscillators, i.e., the microscopic motions,
in the regime of desynchronization
of the phase-locking state, which have not yet been well investigated by the
previous works. Several novel features of this system are found. First,
we find simultaneous phase slips of all oscillators at the onset of
desynchronization of the phase locking state, and quantized phase shifts
in these slips are observed. Moreover, we find the interesting cascade
behavior of coupling-induced chaos and a nice tree structure of transitions
from qausiperiodicity to chaos to qausiperiodicity and periodicity. Then
rich behaviors of synchronizations between chaotic oscillators and chaotic
clusters, which have attracted much attention recently for the coupled Rossler
and Lorenz oscillators, can be also identified for the rather old as well as
popular system (1), of which the individual oscillators have simple periodic
motions without coupling. These findings greatly enlarge the application
perspectives of chaos synchronization. These features of phase dynamics 
are expected to be observable in practical systems by experiments, such 
as coupled laser arrays, Josephson junction chains and coupled electrical 
circuits.

In Eq. (1) the periodic boundary condition $\theta _{i+N}(t)=\theta _i(t)$ is
applied. Without losing generality we scale $\omega _i$ such that 
\begin{equation}
\label{2}\sum\limits_{i=1}^{N} \omega _i=0.
\end{equation}
It is well known that for a given $N$ and $\{\omega _i,\ i=1,...N\}$,
there is a critical coupling $K=K_c$. For $K>K_c$, phase-locking can be
observed, and then we have $\{\stackrel{.}{\theta }_i=0,\ i=1,...,N\}$ , and
each $\theta _i$ is locked to a fixed value. For $K<K_c$, no phase-locking
exists, and $\stackrel{.}{\theta }_i(t)$ are no longer zero. In [8-10], it
is found that if we define an average frequency as 
\begin{equation}
\label{3}\stackrel{\_}{\omega }_i=\lim_{T \rightarrow \infty }%
\frac 1T\int_0^T\stackrel{.}{\theta }_i(t)dt,
\end{equation}
synchronization between different oscillators, in the sense of $\stackrel{\_%
}{\omega }_i=\stackrel{\_}{\omega }_j$, $i\neq j$, can be observed in the
region where strict phase-locking of $\stackrel{.}{\theta }_i=0$ is broken.
It is interesting to investigate how the various
oscillators are led to complete synchronization (phase-locking for $K>K_c$)
via a sequence of bifurcations by increasing the coupling $K$ from $K=0$. 

In order to get a general idea about the global behavior of the system, we
first measure the following two positive macroscopic quantities $R$ and
$\Omega $: 
\begin{equation}
\label{4}
R= \frac 1N\left| \sum
\limits_{j=1}^{N}
e^{i\theta_j}\right| ,
\quad 
\Omega =\lim_{T\rightarrow \infty }\frac 1T
\int_0^T {\sum\limits_{i=1}^{N} }\left| \stackrel{.}{\theta }_i(t)\right| dt.
\end{equation}
It is clear that $R$ is time-dependent beyond the phase-locking region. 
Then we further measure its time average as an order parameter:
\begin{equation}
\label{5}
\left< R \right> =\lim_{T\rightarrow \infty}\frac 1T
\int_0^T R(t)dt,
\end{equation}
In Figs. 1(a) and (c), we plot $\Omega $ vs. $K$ for $N=5$ and $15$,
respectively. In both cases, natural frequencies are randomly chosen from
a normal Gaussian distribution. The actual frequencies can be seen in 
Fig. 3(a) for $N=5$ and Fig. 3(b) for $N=15$ at $K=0$, respectively. These 
natural frequencies are
used for all figures in this letter. We found, for the given natural 
frequencies, $K_c=5.08$ for $N=5$ and $K_c=6.22$ for $N=15$. 
When $K>K_c$ we have identically $\Omega =0$, and then
complete synchronization (phase-locking) is justified. In Figs.1(b) and (d)
we do the same as (a) and (c), respectively, with the measured quantity
replaced by $<R>$. In computing Figs.1, initial conditions of $\theta_j(0)$
are randomly chosen, and then in the shaded region of Fig. 1(d) the
coexistence of multiple attractors of phase-locking states is justified for 
$K>K_c$. In Fig.1, the quantity $T$ in Eqs. (4) and (5) is taken 
sufficiently long in our simulations so that the fluctuations due to 
finite $T$ are invisible. It is interesting to observe that several 
discontinuities
appear in the region $K<K_c$, indicating that, apart from the apparent
phase-locking transition, some additional transitions exist even before
$K_c$.

For the microscopic quantities, it is natural to study the velocities of
various oscillators. In Figs. 2(a) and (b), we present the motions of
$\stackrel{.}{\theta }_i$ vs. $t$ for $N=5$ and different $K$'s. For $K=0$,
$\stackrel{.}{\theta }_i$ must be equal to the constant $\omega _i$, and for 
small
$K$ [see (a)] $\stackrel{.}{\theta }_i$ varies oscillatorily around its natural
frequency. As $K$ increases, the oscillations of $\stackrel{.}{\theta }_i$
become large, and the oscillation centers of all oscillators shift close to
each other. An interesting feature is that near the onset of synchronization
[see (b)], we can find simultaneous phase slips, i.e., all oscillators 
keep in a
the phase-locking condition (OFF-state) for a long time, and then simultaneous
bursts of all oscillators (ON-state) break the locking state. After a short
firings all oscillators calm down again simultaneously to another
phase-locking state, and
then repeat the same process periodically. As $K$ gets closer to $K_c$, the
length of the OFF state $\tau $ becomes longer. We find a clear scaling
between $\tau $ and $K_c-K$:
\begin{equation}
\label{6}\tau \propto (K_c-K)^{-0.5}.
\end{equation}

The above features of synchronized actions of phase slips can be well
understood by an
intuitive explanation. Suppose various oscillators can be locked to a set of
phases $\stackrel{\_}{\theta }_i(K)$ for $K\geq K_c$. Then all solutions
satisfying
\begin{equation}
\label{7}\stackrel{\_}{\theta }_{i+1}(K,{\bf m})-\stackrel{\_}{\theta }_i(K,%
{\bf m})=\stackrel{\_}{\theta }_{i+1}(K)-\stackrel{\_}{\theta }_i(K)+2\pi m_i
\end{equation}
with ${\bf m}=(m_1,...,m_j,...m_N)$ and $m_i$ being any integer, must also
be phase-locking solutions of (1). By reducing $K$ lower than $K_c$, all the
above solutions lose their stability via the saddle-node bifurcation. At $K=K_c$,
there exists a heteroclinic path linking some of the above solutions,
which has the lowest potential, and is attracting (the existence of such a
heteroclinic path can be rigorously proven for $N=2$). For $K<K_c$, and
$\left| K-K_c\right| \ll 1$, the system takes such a periodic path, which
stays in the vicinity of one of the above stationary solution for a long
time, and escapes away from this solution (simultaneous phase slips for all
oscillators), and then approaches to the vicinity of the next stationary
solution alone the heteroclinic path of $K=K_c$; this produces the periodic
pulses of Fig. 2(b). The scaling property of time length of the OFF-state
$\tau $ can be also computed since for the saddle-node bifurcation we have a
universal form $\stackrel{.}{x}=(K_c-K)+x^2.$ The time $\tau $ for $x$ to
move from $x=0$ to $x\rightarrow \infty $ reads $\tau \propto \int_{0}^{\infty} 
\frac{dx}{(K_c-K)+x^2}=\frac \pi {2\sqrt{K_c-K}}$; this explains our
observation of the scaling law of Eq. (6).

It is interesting to compute the phase shifts for various oscillators in
each phase slip pulse in Figs. 2(b). Let $\Delta \theta _i$ represent the phase
shift of $\theta _i$ during each pulse. From (1) and (2) we have 
\begin{equation}
\label{8}{\sum\limits_{i=1}^{N} \Delta \theta _i=0.}
\end{equation}
We argue that any two adjacent fixed points in a heteroclinic path at $K=K_c$
take $m_i=0$ or $\pm 1$ in Eqs. (7), i.e., 
\begin{equation}
\label{9}\Delta \theta _{i+1}-\Delta \theta _i=0\text{ or }\pm 2\pi 
\end{equation}
for $i=1,...,N-1$. Considering both conditions (8) and (9), $\Delta
\theta _i$ can take only quantized values 
\begin{equation}
\label{10}\Delta \theta _i=0,\pm \frac{2\pi }N,\pm \frac{4\pi 
}N,...,\pm \frac{ 2(N-1)\pi }N,\pm 2\pi .
\end{equation}
The concrete value for each $\Delta \theta _i$ depends on the particular
distribution of $\omega _i$. In Figs.2(c) and (d) the phase shifts observed
in all pulses fully agree with our heuristic argument. Further considering Fig.3,
the phase shifts in Fig.2 can be exactly predicted (see [15]). One more
conclusion from the above analysis is that at the onset of instability of the
phase-locking state, all velocities $\stackrel{\_}{\omega }_j$ have the
scaling $\stackrel{\_}{\omega }_j\propto \sqrt{K_c-K}$, and the ratios
between different $\stackrel{\_}{\omega }_j$ must be quantized to
discrete rational numbers.

As we further reduce $K$ to values considerably smaller than $K_c$, no more
phase-locking exists, and no apparent synchronization can be observed directly
for $\stackrel{.}{\theta }_j(t)$. However, some other implicit
synchronization --- phase synchronization --- can be still found. To get a
general idea, we plot $\stackrel{\_}{\omega }_i$ defined in (3) vs. $K$ for
$N=5$ and $15$ in Fig. 3(a) and (b), respectively, by varying $K$ from $K=0$
to $K>K_c$. Interesting behavior of transition tree for phase
synchronization is clearly shown. Though some synchronizations can be
expected from the observations of frequency plateaus reported in previous
papers [8-10], a number of characteristic features revealed in these trees
are novel and interesting. Three kinds of transitions can be observed in
these trees. First, if two {\it adjacent} oscillators (or adjacent clusters of
oscillators) have close frequencies, they can be easily synchronized by
increasing $K$. In this case, one always finds two branches merging to a
single one (indicated by {\bf A}). Second, if two {\it non-adjacent} oscillators
(or two non-adjacent clusters) have close frequencies while the oscillators 
between them
have considerably different frequencies, one can find the non-adjacent
oscillators can be also synchronized to each other, i.e., non-local clusters
can be formed, and these non-local clusters can quickly bring the
oscillators between them to the synchronized status, and form a solidly larger
synchronized cluster. In this case, the transition may be from two to one or
from multiple
branches to one (indicated by {\bf B}). An oscillator, which is synchronized
to a cluster for certain $K$, can be {\it desynchronized} from the original cluster
by increasing $K$. This desynchronization always happens at an edge
oscillator of a cluster, due to the competition between two neighbor
clusters (indicated by {\bf C}, see 2nd and 3rd oscillators of Fig. 3(b)).
It is obvious that {\bf C} is the inverse of {\bf A}. The
transitions of types {\bf B} and {\bf C} have never been realized before.

The most interesting and novel finding is the nature of these synchronizations.
The motions in these synchronization trees may be very different. They can be
periodic, quasiperiodic and chaotic. In Fig. 4(a), we take $N=15$ and plot
the largest Lyapunov exponent of the system vs. $K$. In a large interval of
$K$, we find the positive Lyapunov exponent, indicating chaos. Thus, in this
region phase synchronizations of chaotic oscillators are identified.
Recently, the phase synchronization of coupled chaotic systems has attracted
great attention [11]. Here the major difference between
our system and the
previous chaos synchronization is that our oscillators are periodic without
coupling, and chaos is induced by the coupling, while in the latter case the
individual systems are chaotic without coupling. Moreover, we find that
these coupling-induced chaotic motions of different oscillators may have
different levels of synchronizations by varying the coupling, and the cascade
of synchronization forms a tree-like structure of Fig. 3.

In Figs. 4(b) to (f) we plot the maps of $\stackrel{.}{\theta }_1(n)$ to
$\stackrel{.}{\theta }_1(n+1)$, where $\stackrel{.}{\theta }_1(n)$ is the
$\stackrel{.}{\theta }_1$ value at the time $t$ when $\theta _1(t)$ crosses
the angles $2n\pi $ with $n$ being an integer. For $K>K_c$, we have fixed
point solution, and the map is fixed at 
$\stackrel{.}{\theta }_1(n)=\stackrel{.}{\theta }_1(n+1)=0$. For 
$K$ slightly smaller than $K_c$ we
have periodic solution represented by the finite number of dots in Fig. 4(b).
The period $8$ can be easily understood from Fig. 2(d). The period of the total
system is $15\tau $, where $\tau $ is the time interval between the two adjacent
slips, and the change of of $\theta _1$ in $15\tau $ is $-16\pi $. This 
leads to the period-$8$ solution of Fig. 4(b). Two-frequency torus can be 
identified in the three-cluster regime [see Fig. 4(c)]. For very small $K$,
we can find high-dimensional quasiperiodicity [e.g., Fig. 4(f)].
Chaos is prevailing in the $K$ region between Fig. 4(c) and 4(f) [see Fig.
4(d) and (e), and Fig. 4(a)]. Many periodic windows are found in the
quasiperiodic and chaotic regions, which will be investigated in detail in
our forthcoming extended paper.

The entire variation from high-dimensional quasiperiodicity (for very small
$K$) to periodic motion ($K<K_c$, $\left| K-K_c\right| \ll 1$) through
various orders of chaos synchronizations can be vitally seen in Figs. 3 and
4. Starting from the high-dimensional quasiperiodicity for $K\ll 1$, by
increasing $K$ various neighboring oscillators with close frequencies start
to form clusters via phase synchronization, and chaos is induced near the
first synchronization. Then in each cluster, different oscillators perform
different chaotic motions, while having identical winding number. The
winding numbers for different chaotic clusters are different.
When further increasing $K$,
adjacent chaotic clusters can be synchronized to form
larger clusters, until two large clusters are formed when the
motion becomes periodic. This tree picture of the transitions is expected 
to be
the same for general large number of coupled non-identical oscillators, which 
are periodic in bare case.

Since system (1) qualitatively describes practical situations in wide fields,
ranging from physics, chemistry to biology, the findings in this letter are
expected to be of general significance, and they can be used for understanding
the mechanisms of rich collective behaviors of coupled systems. Laser 
arrays, Josephson junction chains and coupled electrical circuits may be 
ideal candidates for experimentally revealing the features explored in 
this letter.

This work is supported in part by the Research Grant Council RGC and the
Hong Kong Baptist University Faculty Research Grant FRG and in part by the
National Natural Science Foundation of China.

\begin{figure}
\epsfxsize=8cm
\epsfbox{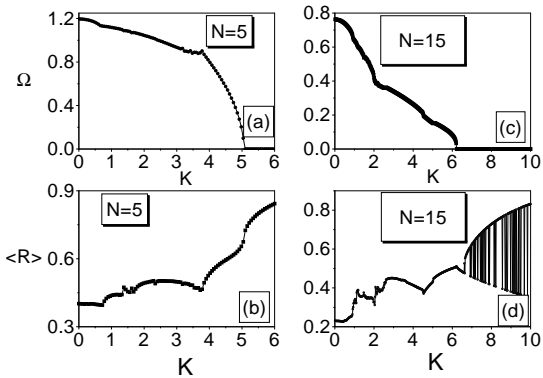}
\narrowtext
\caption{
$\Omega $ and $<R>$ defined in Eqs. (4) and (5) plotted vs. $K$ for $N=5$ and
$15$. The natural frequencies are shown in Fig. 3 at $K=0$. $K_c=5.08$ 
and $6.22$ for $N=15$ and $N=15$. For $K>K_c$, phase-locking can be 
observed, and the shaded region in (d) indicates the coexistence of 
multiple attractors of phase-locking states. Initial conditions are randomly 
chosen for each $K$ detected. } 
\end{figure}

\begin{figure}
\epsfxsize=8cm
\epsfbox{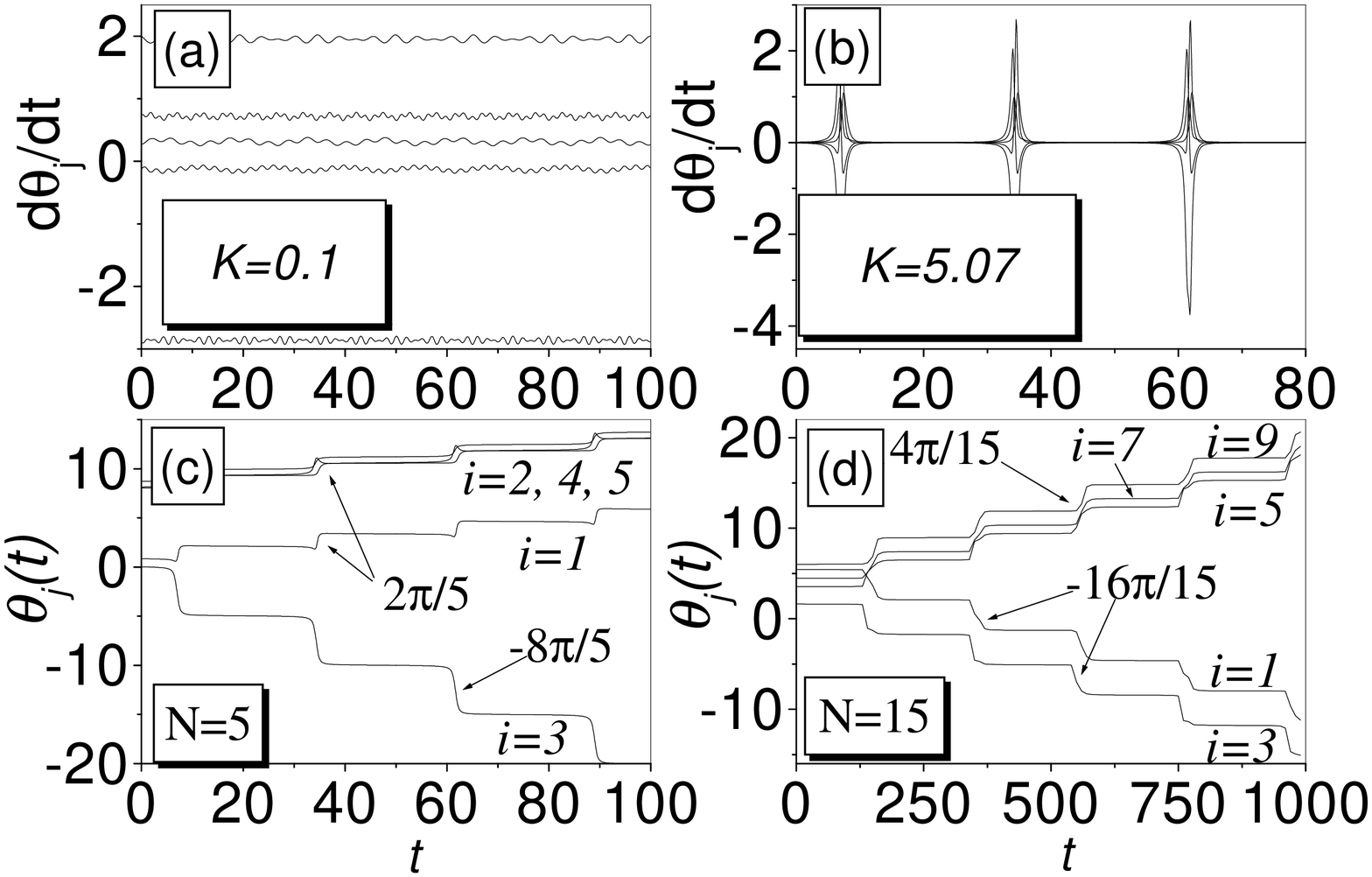}
\narrowtext
\caption{
(a), (b): The time evolutions of $\stackrel{.}{\theta }_i$ plotted for $N=5$
and two different $K$. Near the onset of the instability of the phase-locking
state [(b)], simultaneous phase slips of all oscillators are clearly seen. 
(c), (d): Some $\theta _i(t)$ plotted against $t$ near $K_c$.
(c) $K=5.07$, $N=5$;
(d) $K=6.212$, $N=15$.
Quantization of phase shifts for all oscillators at each phase slip and
the pulse are justified.
}
\end{figure}

\begin{figure}
\epsfxsize=6cm
\epsfbox{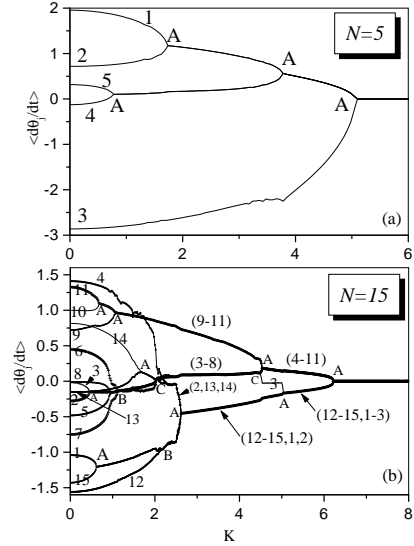}
\narrowtext
\caption{
Transition trees of synchronization for averaged velocities versus $K$.
(a) $N=5$, (b) $N=15.$ Note the existence of three kinds of
transitions labeled by {\bf A}, {\bf B} and {\bf C}.
}
\end{figure}

\begin{figure}
\epsfxsize=8cm
\epsfbox{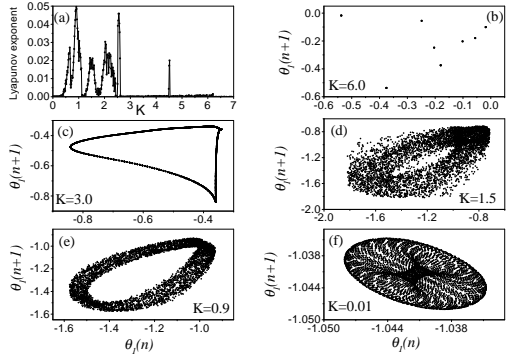}
\narrowtext
\caption{
(a) The largest Lyapunov exponent of the coupled oscillators plotted
against $K$, $N=15$. (b)-(f) give the maps of $\stackrel{.}{\theta }_1(n+1)$
vs. $\stackrel{.}{\theta }_1(n)$ where $\stackrel{.}{\theta }_1(n)$ is the
value of $\stackrel{.}{\theta }_1(t)$ when $\theta _1$ crosses $2\pi n$. (b) 
$K=6.0$, the motion is periodic. (c) $K=3.0$, the motion is quasiperiodic
with two irreducible frequencies. (d), (e): $K=1.5$ and $0.9$, respectively.
The motions are chaotic. (f) $K=0.01$, the motion is high-dimensional
quasiperiodicity.
}
\end{figure}

\end{multicols}

\begin{references}
\bibitem{note} {From Fig.3(a) and Eqs.(7) and (8), we have $\sum\limits_{1,2,4,5}%
\Delta \theta _i=-\Delta \theta _3=4\Delta \theta _i,$ $i=1,2,4,5,$ and
$\Delta $$\theta _2-\Delta \theta _3=2\pi ,$ which lead to $\Delta $$\theta
_3=-\frac{8\pi }5$, and $\Delta $$\theta _i=\frac{2\pi }5$, $i=1,2,4,5$. The
same computation from Fig. 3(b) leads to
$8\Delta \theta _{i=4-11} =7\Delta \theta _{i=12-3}$, $\Delta
\theta_4-\Delta \theta_3=2\pi $, producing  $\Delta \theta _i=\frac{14\pi 
}{15}$ for $i=4-11$ and $\Delta \theta _i=-\frac{16\pi }{15}$ for $i=12-3$.}
\end{references}
\end{document}